\documentclass[runningheads]{llncs}
\usepackage{graphicx}
\usepackage{glossaries}
\usepackage{multicol}

\usepackage{glossaries}
\usepackage{xcolor}
\usepackage[utf8]{inputenc}

\usepackage{listings}
\usepackage{textcomp}
\usepackage{amsmath}  
\usepackage{amssymb} 
\usepackage{url}
\usepackage{booktabs}
\usepackage{listings}
\usepackage{color}
\usepackage{lastpage}

\usepackage{tikz}
\usetikzlibrary{arrows,shadows}
\usepackage{pgf-umlsd}

\usepackage{comment}
\authorrunning{Pasumarthy et al.}

\begin{document}
\title{TIPS: Threat Sharing Information Platform for Enhanced Security}

\author{Lakshmi Rama Kiran Pasumarthy\inst{1} \and Hisham Ali\inst{1} \and William J Buchanan\inst{1} \and Jawad Ahmad\inst{1} \and Audun Josang\inst{2} \and Vasileios Mavroeidis\inst{2} \and Mouad Lemoudden\inst{1}
}

\institute{Blockpass ID Lab, Edinburgh Napier University \email{\{lakshmiramakiran.pasumarthy@napier.ac.uk,h.ali,b.buchanan,j.ahmad\}@napier.ac.uk} \and University of Oslo \email{\{osang,vasileim\}@ifi.uio.no}}

\maketitle              

\begin{abstract}

There is an increasing need to share threat information for the prevention of widespread cyber-attacks. While threat-related information sharing can be conducted through traditional information exchange methods, such as email communications etc., these methods are often weak in terms of their trustworthiness and privacy. Additionally, the absence of a trust infrastructure between different information-sharing domains also poses significant challenges. These challenges include redactment of information, the Right-to-be-forgotten, and access control to the information-sharing elements. These access issues could be related to time bounds, the trusted deletion of data, and the location of accesses. This paper presents an abstraction of a trusted information-sharing process which integrates Attribute-Based Encryption (ABE), Homomorphic Encryption (HE) and Zero Knowledge Proof (ZKP) integrated into a permissioned ledger, specifically Hyperledger Fabric (HLF). It then provides a protocol exchange between two threat-sharing agents that share encrypted messages through a trusted channel. This trusted channel can only be accessed by those trusted in the sharing and could be enabled for each data-sharing element or set up for long-term sharing.

\end{abstract}

\section{Introduction}
With increasing threats in cybersecurity, the requirement to share threat information becomes ever more important. Traditional transport mechanisms such as electronic email often lack any real form of privacy and trust, and so we need improved methods that can identify the sender and receiver of threat-sharing information. Along with this, we need to integrate threat-sharing methods that integrate privacy-aware techniques, especially to comply with data regulations such as the General Data Protection Regulation (GDPR) etc. This should support enhanced features such as the redaction of information, the Right-to-be-forgotten (RTBF), and location/time-based access controls. And, so, end-to-end encryption using messaging applications, such as WhatsApp and Signal can be used for information sharing, the trust infrastructure is generally hosted on external systems, and which limit the usage of digital wallets and in the auditing of the threat sharing exchange. Risks can also exist in external parties \emph{ghosting} onto communications \cite{baraniuk2015ghosts}.
 
The paper provides an abstraction of the information-sharing process between threat agencies and which integrates privacy-aware methods such as Attri-bute-Based Encryption (ABE), Homomorphic Encryption (HE) and Zero Knowledge Proofs (ZKP). It then builds on this to provide a secure and trusted channel for the exchange of information on threats using a permissioned blockchain (Hyperledger Fabric). The main contribution of this paper is to provide a protocol that will allow threat agencies to pass threat information in an encrypted format within a trusted communication channel. A permission blockchain is used as this allows for the integration of trusted threat-sharing agencies, where each threat-sharing agent within a trusted agency can prove their identity and share their public keys in a trusted way. The enactment of this is achieved through the use of chain code.

Maasimo et al. \cite{Massimo} presented the state-of-the-art for Threat information sharing standards and platforms. Several standards have been introduced to facilitate the sharing of threat information, such as Structured Threat Information CybereXpression (STIX), Cyber Observable eXpression (CybOX), Incident Object Description Exchange Format (IODEF), and  Trusted Automated eXchange of Indicator Information (TAXII). Some platforms based on these standards are Malware Information Sharing Platform, Collaborative Research Threats (MITRE CRIT), Collective Intelligence Framework (CIF), EclecticIQ Platform and LookingGlass Cyber. 

Overall, the paper outlines the creation of a trusted communication channel between threat agents where we use digital wallets to identify them and thus digitally sign for trusted information flows. In the case of Alice sending threat-sharing information to Bob, Bob generates his public key pair on his computer  - ($E_{priv}, E_{pub}$). Bob and Alice then set up a trusted Hyperledger channel that cannot be accessed by other threat agents. Bob then places his public key in that channel. Alice can then receive this and encrypt a symmetric key with Bob's public key ($E_{pub}$). She can then place the encrypted threat data and the encrypted key in the channel for Bob to receive. Bob then decrypts this on his computer using his private key. An important advantage of this, is that We can audit and log each part of this process so that Alice knows when Bob has read the encrypted message and in the time that Alice posted it to the secure channel. We can thus make the channel a long-term location for information sharing, or create for each transfer. In this case, RSA public key encryption is used to support the generation of the key pair.

\subsection{Permissioned Blockchain}

A permissioned blockchain, alternatively called a private blockchain, is another type of blockchain network in which accessibility is limited to specific user segments. Hyperledger Fabric is the best example of this kind of blockchain network, whereas Ethereum is an example of a public blockchain. Several industries, such as healthcare, defence, and investigation agencies, can be applications of permissioned blockchains \cite{POLGE2021229}.  Hyperledger Fabric, an open-source framework, is pivotal for building distributed ledgers using blockchain technology, encompassing applications, tools, and libraries \cite{hyperledger-A}.  One significant component is the \emph{Fabric Channel}, providing secure communication among network participants for transactions within a subnet. It ensures trusted data sharing and demonstrates high transaction performance compared to traditional blockchain processes \cite{HyperLedger-fabric}.

\subsection{General Data Protection Regulation (GDPR)}

The General Data Protection Regulation (GDPR), instituted by the European Union in May 2018 \cite{grundstrom2019making}, safeguards the privacy of personal data exchanges among EU member states and affiliated entities globally. Compliance with GDPR is obligatory for all entities engaging in data transactions with EU members, necessitating prior consent. Moreover, GDPR presides over pivotal facets of data authorisation and authentication, encompassing rights such as data access, modification, deletion, and combination, ensuring comprehensive oversight of data-sharing practices.

\subsection {Right-to-be-forgotten (RTBF)}  

The Right to be Forgotten (RTBF) pertains to GDPR mandates concerning data deletion \cite{havelange2019luce}. However, disseminating this right across the network proves impractical and unfeasible. Consequently, the significance of discovering covert protocols arises to facilitate content deletion upon request. Furthermore, this directive may be enforced by content proprietors or their designated representatives.

\subsection {Cyber Threat Intelligence (CTI))} 

Cyber Threat Intelligence (CTI) is an evidence-based knowledge which plays an important role in cyber security. The CTI lifecycle involves the process of data collection, analysis and dissemination among trusted users. Threat intelligence sharing has become crucial, enhancing organisations' security posture against emerging threats and keeping them vigilant to repel potential attacks. To plan for, respond to, and mitigate these risks effectively, the main objective of CTI is to assist organisations in understanding the threats they face, including the tactics, methods, and procedures (TTPs) employed by cyber attackers. Considering the background, intents, and capabilities of threat actors is equally important to CTI in comprehending cyber threats' technical components. However, the most significant challenges encountered by threat intelligence agents are sharing threats across different organisational boundaries, navigating privacy regulations, and implementing suitable technological solutions. To solve these issues, we introduce a novel solution using a unified threat language (STIX) and privacy-preserving methods \cite{ali2022trusted}.

\subsection {Structured Threat Information eXpression (STIX)}

Structured Threat Information Expression (STIX) is a standardised language for describing cyber threat intelligence in a structured manner \cite{ali2022trusted}. It provides a common framework for organisations to share, analyse, and respond to cyber threats effectively. STIX enables the exchange of threat information in a machine-readable format, facilitating automation and interoperability among security systems through different geographical locations. Using STIX, organisations can better understand and mitigate cyber threats, improving their overall cybersecurity posture.

\section{Related Work}

This section provides insights into recent advances at the intersection of privacy regulations, blockchain technology, and threat information sharing.

\subsection{Privacy-preserving Data Sharing}

Coppolino et al. \cite{Sgaglione} proposed an IDS architecture that offers customers the benefits of managed security services without giving third-party entities full access to their sensitive information. The growing number of cyber threats has led organisations and companies to seek managed security services (MSS) to address their information security concerns. However, privacy concerns regarding MSS outsourcing have resulted in a conflict between privacy and security monitoring solutions. This paper proposes an intrusion detection system (IDS) architecture for privacy-preserving security monitoring in MSS using homomorphic Encryption (HE) technology. Homomorphic Encryption enables computations on encrypted data without direct access to decryption, allowing secure processing, storage, and exchange of confidential information. 

José M et al. \cite{DEFUENTES2017127} describe privacy-enhancing Cybersecurity Information Sharing (CIS), which plays a vital role in combating cyber threats and incidents by facilitating improved situational awareness among its members... In the surge of sophisticated cyber attacks, partnerships via CIS are critical to read and respond to future threats. As firms use the cloud as a tool for transmitting data, the risk of data privacy is increasing. It proposed a protocol called PRACIS, a system for CIS networks that ensures the confidentiality of data as it is forwarded and aggregated. This system uses the established STIX standard data format. PRACIS uses traditional format-preserving techniques and homomorphic encryption methods to achieve its goals...
Additionally, after conducting tests on a prototype for a portion of STIX, the outcome displays that up to 689 incidents per minute can be reported, much higher than the typical estimation of 81. Furthermore, the consolidation of 104 incidents can be completed within just 2.1 seconds, and the data transmission overhead is minimal at 13.5~kbps. Overall, these findings suggest that the cost of implementing PRACIS is reasonable and practical for practical use. 

David W.C. et al. \cite{CHADWICK2020710} proposed a collaborative model to address these problems in sharing threat information. Sharing Cyber Threat Information (CTI) can significantly aid in forecasting, preventing, and mitigating cyber-attacks. The person or organisation that owns the Data can choose a suitable level of trust and method for sanitising Cyber Threat Information (CTI) data. This can involve using plain text, anonymisation or pseudo-anonymisation strategies, or homomorphic encryption. Before the CTI data is distributed for analysis, it is processed and secured using these techniques. This paper proceeds with associated research on cloud-edge computing, CTI data sharing, and data security concerns in cloud-edge computing. The proposed trust model is then described, and it goes in-depth into the data-sharing infrastructure and explores the diverse deployment models available. Four real-world projects serving as test cases for the data-sharing infrastructure are explained, including their chosen levels of trust and deployment models to meet their unique trust requirements. An overview of the progress made thus far, the testing methodology was implemented, and information on validation work completed to date was provided. It concludes by discussing the current limitations and identifying areas requiring additional research.

\subsection{Architecture of C3ISP}

Cyber threat intelligence has been proposed to achieve cyber threat information sharing between organisations. Various frameworks, such as C3ISP, have been proposed. C3ISP (Collaborative and Confidential Information Sharing and Analysis for Cyber Protection) is funded by the European Commission. The main goal of C3ISP is to create a framework in which entities can share cyber threat information securely and confidentially. This is designed to allow more collaboration between organisations, industries, and countries to better respond to and prevent cyber threats \cite{C3ISP}.

According to Massimo et al. \cite{FanW}, threat information sharing, particularly in Indicators of Compromise (IoC) and threat events, is challenging, and the entities essential to the process of gathering verified data for machine learning algorithms are not adequately supported by current threat information sharing solutions, making it difficult or impossible to communicate with them. So ORISHA - a platform for ORchestrated Information SHaring and Awareness - is a platform that aims to make it easier for threat detection systems, especially those that use Machine Learning (ML) techniques like intrusion detection systems (IDS), to communicate and share information. By coordinating an interconnected network of Malware Information Sharing Platforms, ORISHA aims to address these issues. A distributed Threat Intelligence Platform is formed by this network, making it possible to communicate with various organisations’ Threat Detection layers. Improved prediction of accuracy in Threat Detection Systems can be improved within this ecosystem. In addition to the above features, the ORISHA platform interfaces with a honey net, which enables adding positive attack instances and information to the knowledge base. It was initially tested on various popular IDS and obtained good results. 

\section{TIPS Infrastructure and Methodology}

This section defines a privacy-aware threat-sharing architecture, which will be built by analysing the use case, which will be implemented, and the experimental work will provide insights into the overall architecture. These will address the key gaps within the existing research work and practice. This work uses Hyperledger Fabric, encryption and STIX Protocol \cite{rantos2020interoperability}, along with using TAXII (Trusted Automated eXchange of Indicator Information) \cite{connolly2014trusted}. With this platform, the research will integrate smart contracts and Hyperledger Fabric channels to provide trusted identity checks and private areas for information sharing. To control access to threat-sharing messages, attribute-based encryption (ABE), which is an access control mechanism for cloud storage, will be used. This provides secure and adaptive sharing encryption-based policies and authentication. Two core attributes include time and location, where a receiver must give a signed attestation of the current time and place for these to be included in the policy.

\begin{figure*}
\begin{center}
\centering
\includegraphics[width=1.0\linewidth]{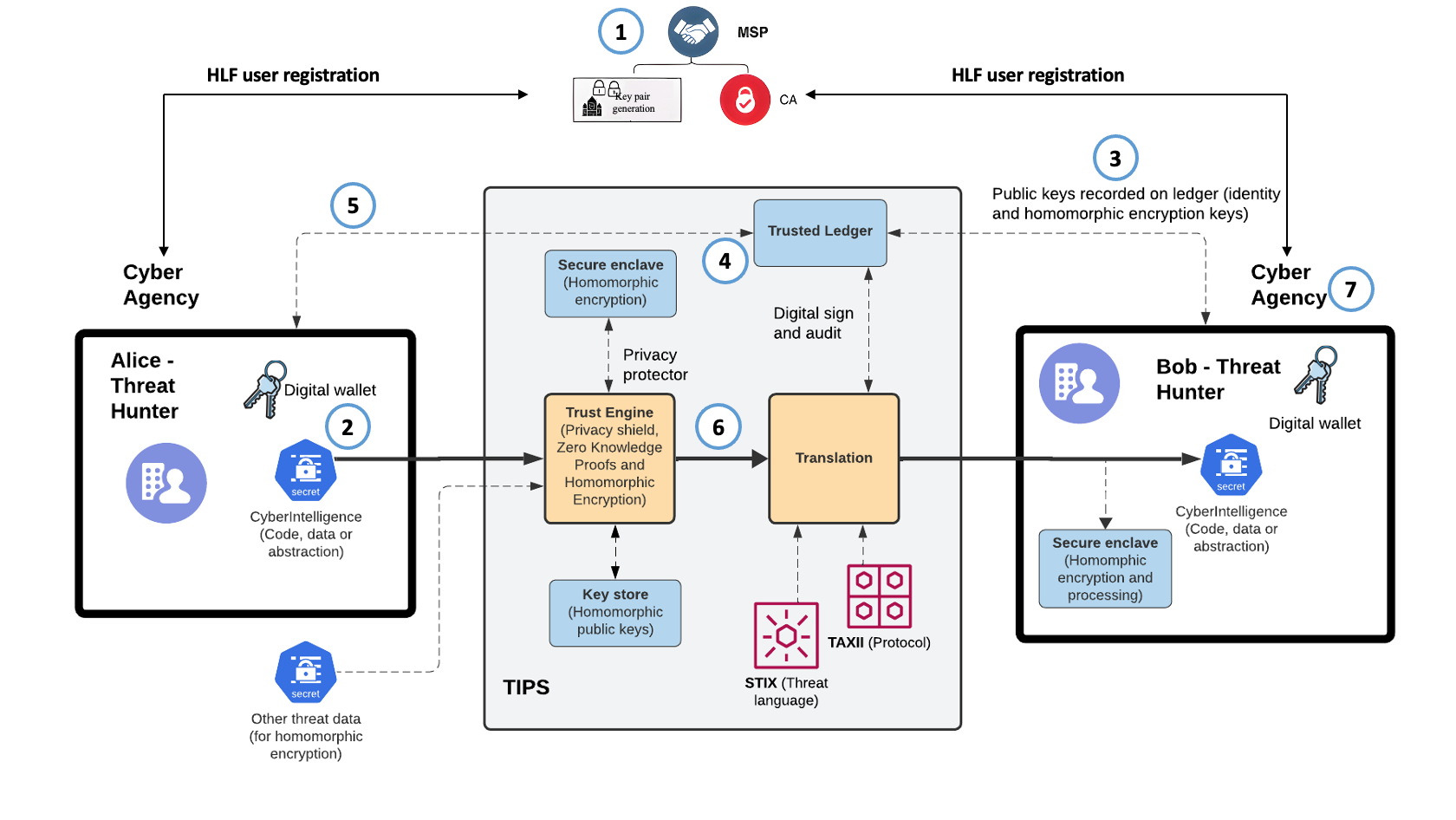}
\caption{Threat information sharing model with ABE, Homomorphic encryption and Zero Knowledge proof}
\label{fig:ThreatSharingModel}
\end{center}
\end{figure*}

Trusted information sharing would happen through a trusted ledger, which will integrate a digital wallet, where Bob and Alice's public keys are stored on the shared ledger. Figure \ref {fig:ThreatSharingModel} outlines the communication between Alice and Bob using the ABE and in supporting interoperability. In ABE, the data is decrypted based on different attributes and policies. 

Based on the literature review and background study, Hyperledger Fabric can be one of the best frameworks for the interoperability and implementation of this model—the design aspects for a sharing data element between Bob and Alice. Alice, who represents a cyber threat agency, is a sender who sends a cyphertext encrypted by the public key of the receiver (Bob); Bob retrieves the message from the trusted cyber threat agency.

In this scenario, the communication happens within a private channel in a ledger, as explained in the steps shown in Figure~\ref{fig:ThreatSharingModel}:

    \begin{enumerate}

        \item MSP manages the lifecycle of user identities, including registration, renewal, and revocation, to maintain network security and integrity.
        Managing user identities in Hyperledger Fabric involves key pair generation, certificate issuance, enrollment, MSP configuration, and ongoing identity management to ensure secure and authenticated interactions within the network.
        
        \item Use Hyperleger to create two wallets.

        \item The ledger is used to store users' public keys, identify Bob and Alice, and support digital signing. There may be the opportunity to blind their identity using a privacy-aware digital signature. The ledger is used to audit shared information.

        \item Setup a trusted channel on Hyperledger.
        
        \item Get the public key of the other person (generated off-chain).

        \item Alice passes the encrypted data (STIX file) to the channel using Bob's public key. 

        \item ABE defines the attributed attributes to read the threat-sharing information. Bob uses its private key to decrypt STIX files that contain threat information. ZKPs are used to verify things without revealing the data source, such as for someone's proof of identity (such as proof of the ownership of a hardware token). Homomorphic encryption is used to aggregate data together from multiple threat-sharing instances.
        
    \end{enumerate}

A well-designed threat-sharing platform has several key attributes which increase its trustworthiness \cite{burger2014taxonomy} \cite{wagner2018novel} \cite{mavroeidis2017cyber}. A firm basis to enhance data security is provided by the permissioned blockchain system Hyperledger Fabric, particularly when it involves threat intelligence information exchange between threat agencies. This is especially important for data security, as timely and safe threat intelligence sharing can greatly strengthen defences against collective threats. Here are a few ways that Hyperledger Fabric can help with this.

To enable the secure communications, Bob and Alice setup a Hyperledger channel for their communication, either for a one-time session, or with a long-term session. Bob will generate a key pair, and where the private key is:

\begin{equation}
k_d\in\mathcal{K}_d
\end{equation} 

and a public key of: 

\begin{equation}
k_e=\mathcal{E}(k_d)\in\mathcal{K}_e    
\end{equation}

This public key is then passed to the Hyperledger channel, and where Alice picks it up. She then generates a random symmetric key of $k_m$ and her threat-sharing message ($m$) of:

\begin{equation}
m\in\mathcal{M}    
\end{equation}

Alice will then encrypt the threat-sharing message with:

\begin{equation}
c=e_{k_m}(m)
\end{equation}

and encrypt the symmetric key with:

\begin{equation}
k_s=e_{k_e}(k_m)
\end{equation}

She will then pass $c$ and $k_s$ into the Hyperledger channel. Bob will pick these up, and then discover the symmetric key using:

\begin{equation}
k_m=d_{k_d}(k_s)
\end{equation}

and then the message with:

\begin{equation}
m=d_{k_m}(c)
\end{equation}

\section{Implementation and Results}

Sharing threat information is vital in cybersecurity battles, yet several key challenges persist. Firstly, there are no clear guidelines for sharing among organisations. Secondly, privacy and legal concerns arise. Thirdly, interoperability issues hinder understanding of different systems' threat data. Lastly, building trust across organisations proves difficult. This work aims to tackle these challenges by proposing strategies to facilitate seamless and trusted threat information sharing, ultimately enhancing online security for all.

This section will provide a practical aspect using Hyperledger Fabric and robust encryption techniques to address these challenges. Specifically, we will utilise a specific environment to install a Hyperledger Fabric network and adapt the encryption techniques.

\subsection{Users Registration on MSP, and Encryption Approach}

The Membership Service Provider (MSP) registration and encryption process plays a pivotal role in achieving the objectives. MSP registration involves authenticating network participants and assigning them cryptographic identities, while encryption ensures the confidentiality of data exchanged within the network. Understanding the intricacies of MSP registration and encryption processes is essential for effectively managing access control and safeguarding sensitive information within Hyperledger Fabric networks. This section delves into the MSP registration and encryption process, elucidating the underlying mechanisms and their significance in bolstering the security posture of Hyperledger Fabric-based blockchain networks.

\begin{figure*}
\begin{center}
    \centering
    \includegraphics[width=1.05\linewidth]{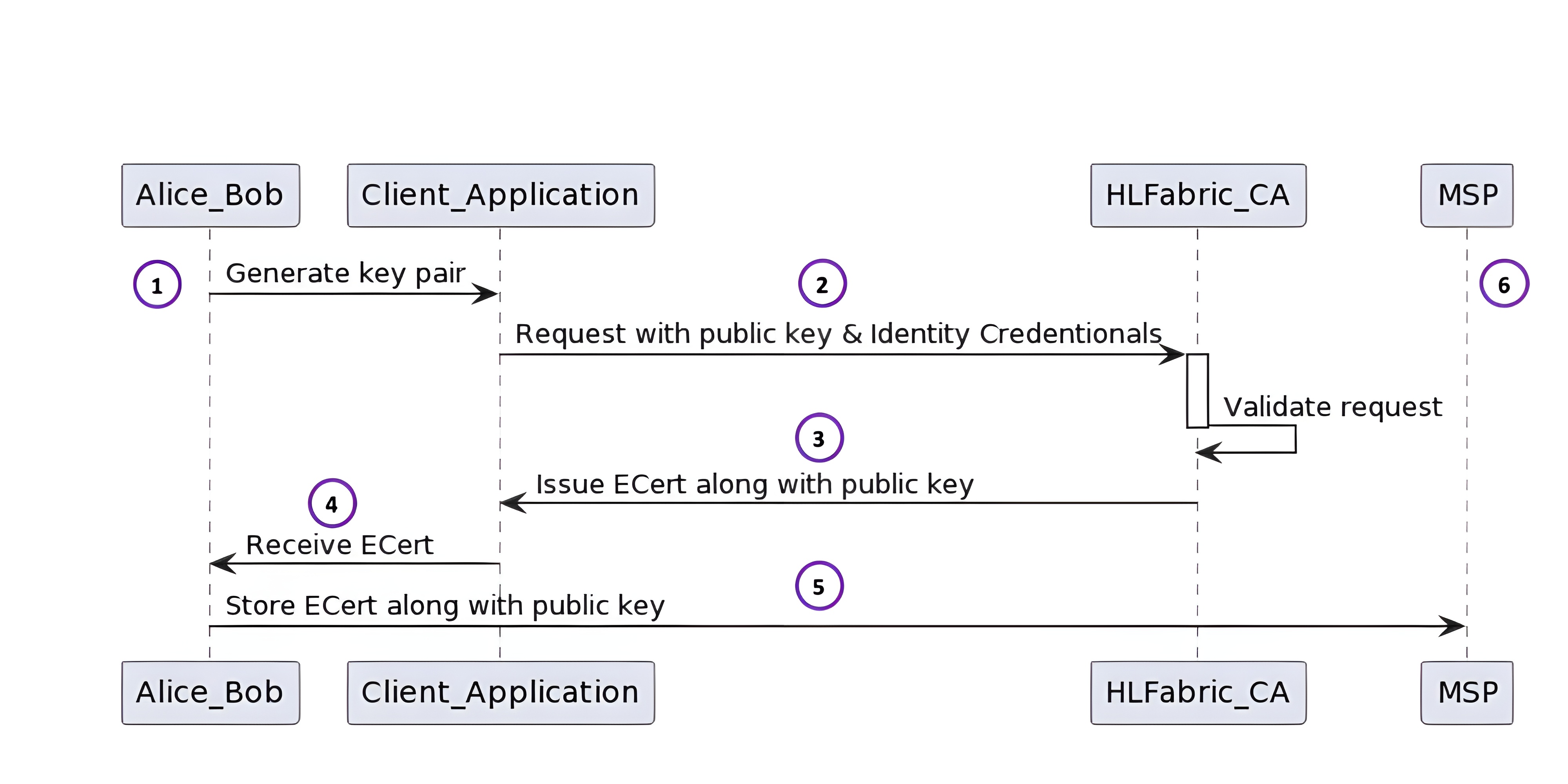}
    \caption{Sequence diagram of Public key management and storage in Hyperledger Fabric}
    \label{fig:Public_Key_storage_Ledger}
\end{center}
\end{figure*}

Here's an overview of the process for managing user identities in HLF, starting from the generation of key pairs as shown in Figure~\ref{fig:Public_Key_storage_Ledger}:

\begin{enumerate}

    \item  Key Pair Generation: Users generate a key pair consisting of a public key and a private key using RSA cryptographic algorithms.  
    
    \item  Certificate Signing Request (CSR): Users create a Certificate Signing Request (CSR) containing their public key and identity information (e.g., Common Name, Organisation).
    The CSR is sent to the Certificate Authority (CA) for certificate issuance.

    \item Certificate Issuance: The CA receives the CSR and verifies the user's identity.
    If the identity is valid, the CA issues an X.509 digital certificate containing the user's public key and identity information.
    
    The CA's private key signs the certificate, making it tamper-proof and verifiable.
    \item  Membership Service Provider (MSP) Configuration: The HLF network defines an MSP for managing identities.
    
    MSP configuration specifies the trusted CA(s), certificate revocation lists (CRLs), and access control policies.
    
    \item  Enrollment: Users enrol in the HLF network by presenting their digital certificate to the Membership Service Provider (MSP).
    
    MSP verifies the certificate's authenticity and validity against the trusted CA(s).
    Upon successful enrollment, users are granted membership status within the network.
    
    \item  Identity Management: MSP manages user identities, including registration, revocation, and attribute-based access control. Users are assigned roles and permissions based on their identity attributes (e.g., organisation, role).

\end{enumerate}

Overall, managing user identities in Hyperledger Fabric involves key pair generation, certificate issuance, enrollment, MSP configuration, and ongoing identity management to ensure secure and authenticated interactions within the network.

\subsection{Messages Exchange Phases}

Message transaction processes through HLF, enabling secure and transparent data exchange among network participants. This process includes transaction proposal, endorsement, ordering, validation, and commitment, ensuring the integrity of transactions on the ledger. Understanding this process is essential for grasping the mechanisms governing blockchain operations in HLF. This section explores the message transaction processing in HLF, outlining its key components and workflow for executing and validating transactions within a Hyperledger Fabric-based blockchain network as shown in Figure~\ref{fig:mt} explained in the following sequences:

\begin{figure*}
\begin{center}
    \centering
    \includegraphics[width=1.03\linewidth]{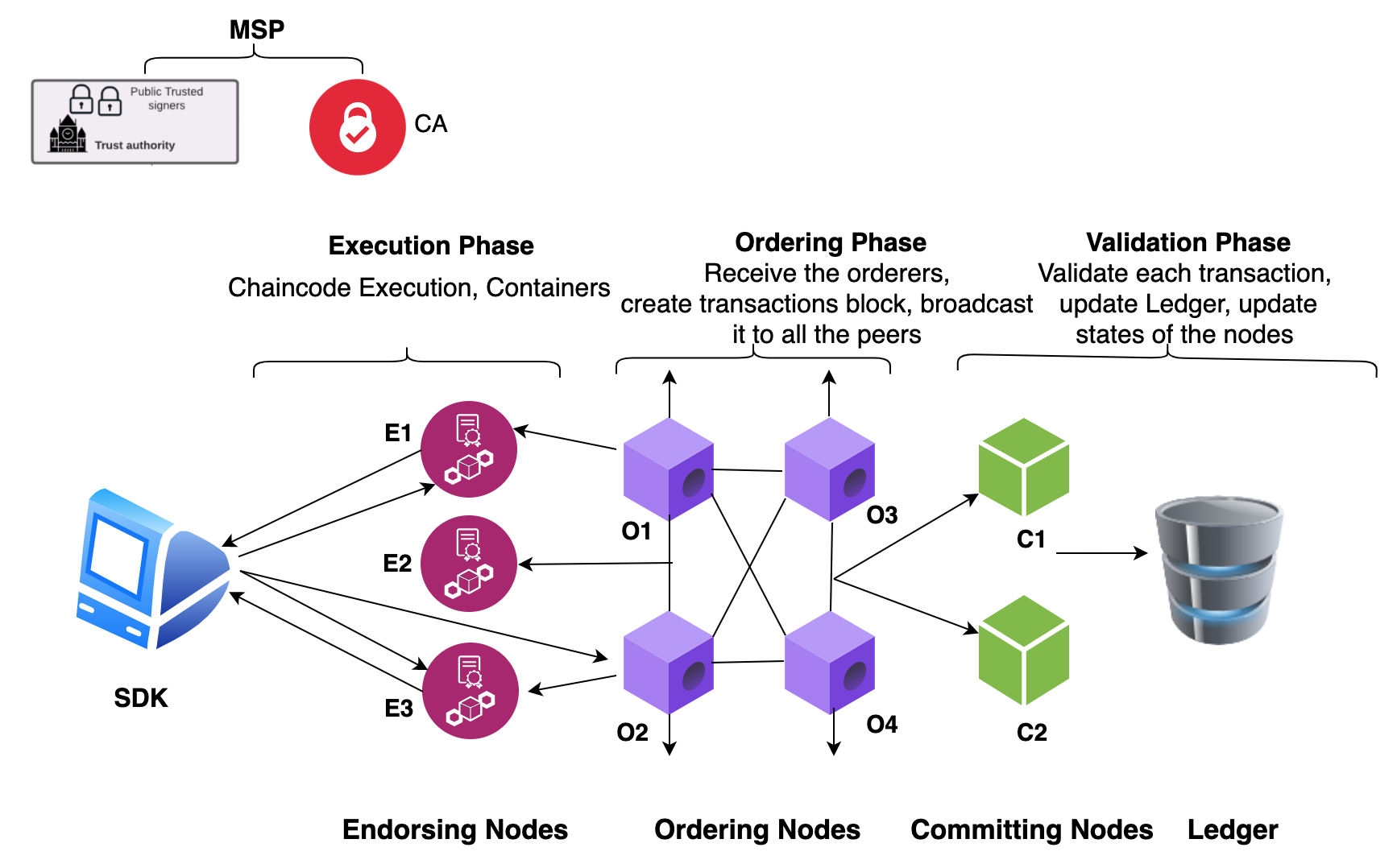}
    \caption{Sequence diagram of Message Transaction within Hyperledger Fabric \cite{ali2021privacy}}
    \label{fig:mt}
\end{center}
\end{figure*}

\begin{itemize}

    \item  Peer and Client Configuration:  Peers and clients in the HLF network are configured with MSP  settings, including the trusted CA(s) and identity information.
     MSP ensures that only authorised entities can participate in network activities.

    \item Transaction Endorsement and Verification: During transaction endorsement, peers verify the identity and signature of the transaction proposer using their digital certificate.  MSP enforces access control policies to ensure that only authorised users can endorse transactions.

    \item  Consensus and Commitment: Peers reach a consensus on transaction validity based on endorsement policies and transaction content.  MSP verifies the integrity of transaction endorsements and ensures that only valid transactions are committed to the ledger.

    \item  Identity Revocation: If a user's privileges need to be revoked (e.g., due to departure from the network or policy violation), MSP updates the Certificate Revocation List (CRL) to invalidate the user's certificate.

    \item  Identity Lifecycle Management: MSP manages the lifecycle of user identities, including registration, renewal, and revocation, to maintain network security and integrity.

\end{itemize}

\subsection{Environment Setup and Test Configurations}

This section includes network performance and functionality tests. Initially, fabric binaries are installed via a command line interface to establish a two-organisation network, with each organisation having two peers linked through a private channel. Subsequently, the benchmark engine engages with chaincode to deploy, execute, analyse, and produce network performance reports, depicted in Fig~\ref{figure:runcaliper}. The setup of the Hyperledger Fabric network requires specific prerequisites, detailed in Table~\ref{tab:my-table3} includes the installation of Hyperledger Fabric binaries package version 2.2.0; Ubuntu version 20.04; Docker version 20.10.7; and Docker Compose version 1.25.0 for generating and managing network entities. Hyperledger Fabric offers three optional languages: Golang, Java, and JavaScript.  Here, we utilised JavaScript (Node.js version 10.19.0) to build chaincode (smart contract).

\begin{table*}
\centering
\caption{Hyperledger Fabric environment setup}
\label{tab:my-table3}
\begin{tabular}{|l|l|}
\hline

{System \& Tools} & {Version} \\ \hline

Operating System         & Ubuntu 20.04     \\ \hline

Hyperledger Fabric       & 2.2.0            \\ \hline

Docker                   & 20.10.7          \\ \hline

Docker-compose           & 1.25.0           \\ \hline

Node.js                  & 10.19.0          \\ \hline
\end{tabular}
\end{table*}

\subsection {Interaction System Implementation}

The implementation extends the asset-transferbasic/ chaincode-javascript using a Solo Ordering Service provided by Hyperledger Fabric and a test network. Performance testing of the smart contract on a Fabric network is conducted using Caliper. The basic workflow involves implementing smart contracts using Hyperledger Fabric, deploying them with JavaScript, and testing by installation, approval, commitment to the channel, and invocation of chaincode \cite{ali2023passion}. 

The chaincode handles various data queries and addresses constraints of threat information sharing by using hash keys for lightweight implementation on endorsing peers. The system's chaincode operations include storing data, querying checksums, retrieving objects, extracting versions, retrieving lineage, and more.

\begin{figure*}
\centering
\includegraphics[width=1\linewidth]{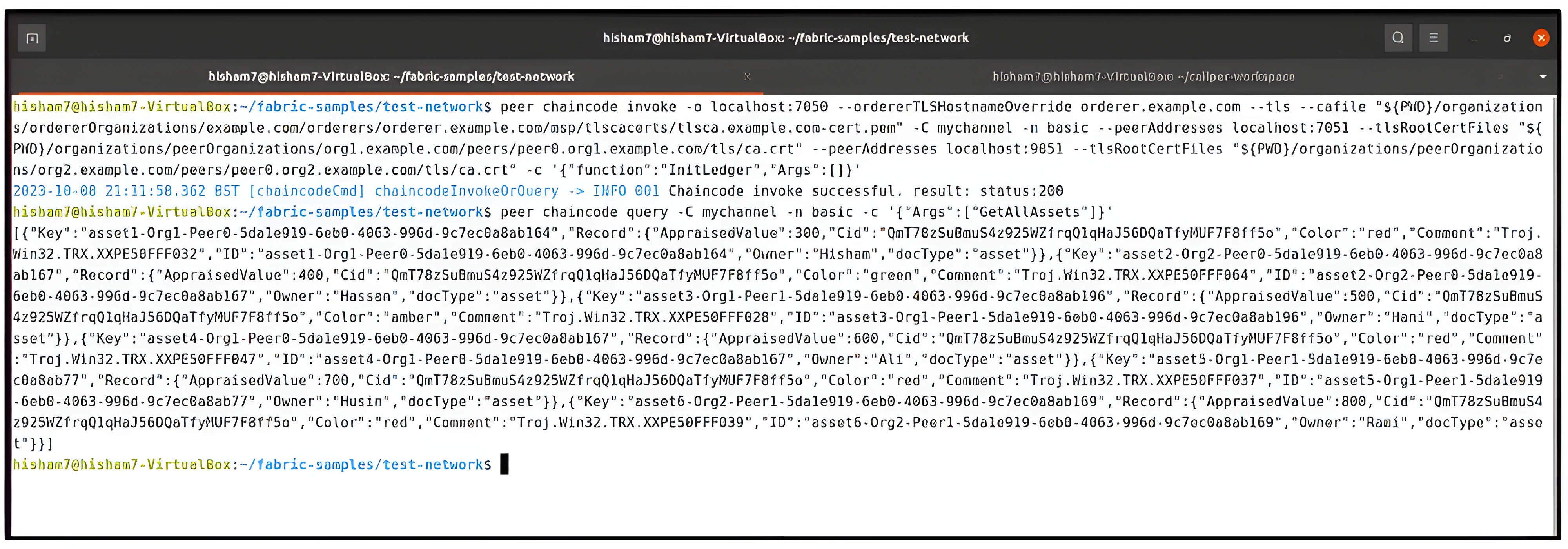}
\caption{chaincode invocation with majority endorsement peers}
\label{figure:chaincodeinvocation}
\end{figure*}

\begin{figure*}
\centering
\includegraphics[width=1.07\linewidth]{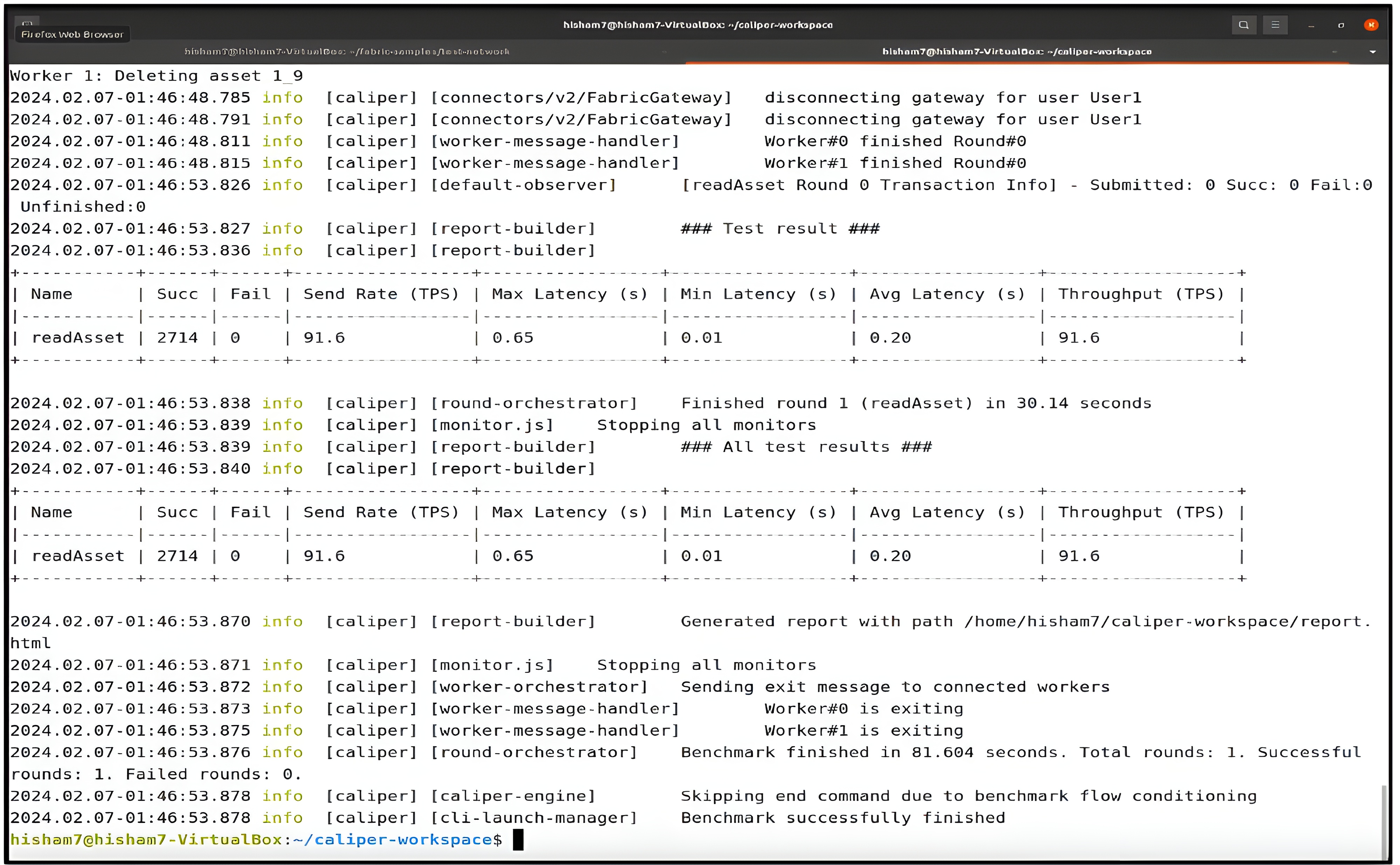}
\caption{Running Caliper benchmark and getting the network performance report}
\label{figure:runcaliper}
\end{figure*}

Performance testing encompasses system throughput, latency, scalability, and resource usage, evaluated with different transaction loads and asset batch sizes. Benchmarks involve 'getAssetsFromBatch' transactions, with evaluations conducted using Caliper benchmarking tools. The assessment includes real-time data reporting and resource consumption statistics for Hyperledger Fabric V2.2.0. The subsequent steps below illustrate different functions within Hyperledger infrastructure setup and network performance assessment:

\begin{itemize}

\item  Start the test network and establish the channel.
\item Package and deploy the smart contract.
\item Approve the chaincode definition.
\item  Commit the chaincode definition on the channel.
\item Execute the chaincode, as depicted in Figure~\ref{figure:chaincodeinvocation}.
\item Integrates benchmarking tools to assess network performance, monitoring network latency, send rate, and throughput, as demonstrated in Figure~\ref{figure:runcaliper}.
\end{itemize}

\subsection {Results}
 
 Performance benchmarking was conducted using the Hyperledger Caliper Benchmarking Tool, focusing on fundamental performance indicators during the message exchange between Alice and Bob. Our results show reading measurements, achieving approximately 91.6 transactions per second. During extensive testing, we observed a maximum latency of 0.65~seconds, a minimum latency of 0.01~seconds, and an average latency of 0.20~seconds. The latency is the time for the ciphertext and encrypted session key to be committed to the ledger and then to be read by the recipient.
 
 These results of scalability, throughput and latency are derived from Hyperledger Fabric's Execute-Order-Validate method, which separates transaction execution and ordering. This separation enhances scalability and performance and reduces node workload. Unlike traditional blockchain architectures, Fabric's approach enables parallel transaction processing, mitigating smart contract non-determinism and yielding higher throughput with lower latency. Consequently, it fosters an efficient and high-performance blockchain ecosystem for trusted sharing.

\vspace{-0.3cm}
\section{Conclusions and Future Work}

Based on the performance benchmarking results and an in-depth understanding of message transaction processes through HLF, this research highlights the importance of privacy-preserving threat information sharing.  The method of using RSA encryption to pass the threat-sharing information provides a mechanism to allow the recipient to pass their public key within a trusted channel and for the private key to be stored in a secure enclave or on the recipient's computer. This channel can be set up either for each session or for a long-term session. While acceptable when a secure enclave is used, it has weaknesses related to \emph{forward secrecy} in that the recipient's private key is breached, especially in creating a long-term channel. For this, an enhancement would look towards PAKE (Password-Authenticated Key Exchange) methods, such as OPAQUE \cite{jarecki2018opaque} to allow for a symmetric key to be exchanged within an asynchronous hand-shake of the key. 

\vspace{-0.5cm}
\bibliographystyle{IEEEtran}
\bibliography{ref}

\end{document}